\documentclass[aps,prb,twocolumn,a4paper]{revtex4}
\usepackage{graphicx}
\usepackage{amsmath}
\usepackage{mathrsfs}
\usepackage{bm}
\usepackage[usenames,dvipsnames]{color}
\usepackage[normalem]{ulem}
\newcommand{\comment}[1]{\textcolor{red}{#1}}
\renewcommand{\comment}[1]{\relax}
\newcommand{\tobedeleted}[1]{\textcolor{green}{\sout{#1}}}
\renewcommand{\tobedeleted}[1]{\relax}
\newcommand{\newtext}[1]{\textcolor{blue}{#1}}
\renewcommand{\newtext}[1]{#1}

\begin{document}

\title{A method for atomistic spin dynamics simulations: implementation and examples}

\author{B. Skubic}
\author{J. Hellsvik}
\author{L. Nordstr\"om}
\author{O. Eriksson}
\affiliation{Department of Physics and Material Science, Uppsala University, Box 530, SE-751 21 Uppsala, Sweden}

\date{\today}
\begin{abstract}
We present a method for performing atomistic spin dynamic simulations. A comprehensive summary of all pertinent details for performing the simulations such as equations of motions, models for including temperature, methods of extracting data and numerical schemes for performing the simulations is given. The method can be applied in a first principles mode, where all interatomic exchange is calculated self-consistently, or it can be applied with frozen parameters estimated from experiments or calculated for a fixed spin-configuration. Areas of potential applications to different magnetic questions are also discussed. The method is finally applied to one situation where the macrospin model breaks down; magnetic switching in ultra strong fields.
\end{abstract}
\maketitle

\section{Introduction}
With the increasing interest in advanced magnetic materials for data storage and processing there is an increasing need for a detailed atomistic description of magnetic materials. Methodological and computational schemes for performing atomistic magnetization dynamics have been presented by several groups in the past.\cite{Ujfalussy1999,Nowak2005,Fahnle2005} At this stage however, there has not been much simulations on realistic systems in a materials research scope. Part of the reason is the computational complexity of these simulations. This limitation is however gradually being overcome by the increasing availability of computational power. At this stage approximate simulations of realistic systems are already feasible and in the future an increasing importance of atomistic modeling of magnetization dynamics can be expected. With recent developments in experimental techniques for studying magnetization dynamics on short time scales and with recent findings on ultrafast magnetization dynamics\cite{Rasing,Stohr}, there is also an increasing amount of experimental results on microscopic magnetization dynamics.

The commonly used approach for studying magnetization dynamics, micromagnetism, provides a framework for understanding magnetization dynamics on length scales of micrometers and has with increasing computational power become a field of large technological importance. The approach, however, suffers from a number of limitations. It is based on the phenomenological Landau-Lifshitz-Gilbert (LLG) equation where magnetism is treated as a continuum vector field on a length scale of micrometers and where energy dissipation from the system is described in terms of a single \emph{ad hoc} damping parameter. This foundation limits the applicability and accuracy of the approach making it inadequate for describing various modern experiments on magnetization dynamics. Instead it would be desirable with an atomistic approach based on the quantum description of solids, an approach which properly displays the connection between the electronic structure of the material and the magnetization dynamics. Such an atomistic approach would be capable of giving a much more accurate description of magnetization dynamics and would provide a framework for including a detailed description of the different dissipation processes involved in magnetization dynamics. It would provide a way of calculating magnetization dynamics starting from first-principles enabling the study of dynamics of materials with complex chemical composition and materials with complex magnetic ordering such as anti-ferromagnets, spin-spirals and spin-glasses.

A formal platform with which to develop an ab-initio spin-dynamics simulation method is naturally based on density functional theory, since it is known to reproduce both magnetic moments as well as exchange interactions with good accuracy. In this paper we have indeed utilized the efficiency of density functional theory in calculating interatomic exchange interactions. The method presented here is based on a Born-Oppenheimer like approximation for the spin system, where we consider the atomistic spins as being slow variables, and the electronic motion being very fast. With this \newtext{adiabatic} approximation one can separate the spin system from the electronic one, as shown by Antropov \textit{et al.}~\cite{Antropov1996}, and hence solve the equations of motion for the two systems separately. 
This is the approach we will adopt here but it is worth mentioning that alternative computational schemes for spin dynamics on an electronic level, is  the time-dependent spin density functional theory (TD-SDFT)\cite{Qian2002} or time-dependent current density functional theory (TD-CDFT)\cite{Vignale1996}. \tobedeleted{Although}These approaches are promising but they are computationally much too time consuming for simulating larger systems. 

The scope of this article is to give a detailed presentation of a methodological and computational scheme for performing spin dynamic simulations on an atomistic scale, where most of the conceptual details were derived in Ref.~\onlinecite{Antropov1996}. The approach is hence based on an atomic scale description of the magnetization of a solid. Magnetic properties extracted from such a description have long been limited to ground state properties as in density functional theory (DFT) or to thermal equilibrium properties as accessed by a combination of DFT and Monte Carlo (MC) simulations. In Section II the adiabatic equations of motion for the atomic spins are derived and here we also discuss magnetic relaxation. In Section~\ref{sec:temperature} we present a scheme for simulating finite temperatures in spin dynamics. Section~\ref{sec:extracting} presents methods of extracting and comprehending results from magnetization dynamics simulations. Finally in Section~\ref{sec:switching}, as a demonstration, the method is applied to magnetic switching of bcc Fe in ultra strong switching fields.

\section{Equations of motion}
\subsection{Slow variables}
A detailed derivation of how the dynamics of fast variables (electrons) and slow variables (atomic spins) is separated can, as mentioned, be found in Ref.~\onlinecite{Antropov1996}. Here we give a short description of the essential aspects of the dynamics of the atomic spins.

The derivation is based on the Kohn-Sham (KS) Hamiltonian ($H_{KS}$) of density functional theory,\cite{kohn} which can be expressed in terms of a non-magnetic ($\mathscr{H}_{\textrm{nm}}$) and a magnetic part ($\mathbf{B}$) as
\begin{equation}
\mathscr{H}_{\textrm{KS}}=\mathscr{H}_{\textrm{nm}}+\mathbf{\hat{\sigma}}\cdot\mathbf{B}.
\end{equation}
The equation of motion for the slow variables, or the directions of the atomic spins, can be derived by evaluating the commutator between the spin operator, $\mathbf{\hat{S}}$, and the Kohn-Sham Hamiltonian 
\begin{equation}
\frac{\partial \mathbf{\hat{S}}}{\partial t}=\frac{1}{i\hbar}[\mathbf{\hat{S}},\mathscr{H}_{\textrm{KS}}],
\end{equation}
which results in
\comment{parentheses}
\begin{equation}
\frac{\partial \mathbf{\hat{S}}}{\partial t}=-\gamma \mathbf{\hat{S}} \times \mathbf{B} +\frac{1}{i\hbar}[\mathbf{\hat{S}},\mathscr{H}_{\textrm{nm}}].
\label{eq:conteq}
\end{equation}
In absence of spin-orbit coupling, $\mathbf{\hat{S}}$ commutes with all terms of $\mathscr{H}_{\textrm{nm}}$ except for the kinetic term $-\frac{\hbar^2}{2m}\sum_i^N\nabla_{\mathbf{r}_i}^2$ (see Ref.~\onlinecite{Capelle2001}). We define the current operator as
\begin{equation}
\mathbf{\hat{j}} \equiv \frac{\hbar}{i2m} \sum_{i}^N \Big(\nabla_i \delta(\mathbf{r}-\mathbf{r}_i) + \delta(\mathbf{r}-\mathbf{r}_i) \nabla_i\Big),
\label{eq:currentdef}
\end{equation}
and the spin-current operator as \comment{not a sum again ...? Q is used for quadrupole later on, maybe J, I have changed accordingly ...}
\begin{equation}
\mathbf{\hat{J}} \equiv  \mathbf{\hat{\sigma}} \otimes \mathbf{\hat{j}},
\label{eq:spincurrentdef}
\end{equation}
where summation in Eq.~(\ref{eq:currentdef}) is performed over electrons. Evaluating the last term in Eq.~(\ref{eq:conteq}) using the stated definitions results in,
\begin{equation}
\frac{1}{i\hbar}[\mathbf{\hat{S}},\mathscr{H}_{\textrm{nm}}]=\frac{1}{i\hbar}[\mathbf{\hat{S}},\Big(-\sum_i^N\frac{\hbar^2\nabla_{\mathbf{r}_i}^2}{2m}\Big)]=\nabla \cdot \mathbf{\hat{J}},
\end{equation}
and the continuity equation for the spin magnetization within the KS framework is obtained by inserting this result in Eq.~(\ref{eq:conteq}). By calculating the expectation value of $\frac{d\mathbf{\hat{S}}}{\partial t}$ for the KS ground state we obtain,
\begin{equation}
\frac{\partial \mathbf{S}}{\partial t}(\mathbf{r},t)+\nabla\cdot \mathbf{J}(\mathbf{r},t)=-\gamma \mathbf{S}(\mathbf{r},t) \times \mathbf{B}(\mathbf{r},t).
\label{eq:KSconteq}
\end{equation}
The second term on the left hand side is omitted for the applications considered in this paper. Among effects that arise from this term are fluctuations of the size of the atomic spins. For experiments when current induced effects are important one must include this term. By using the atomic moment approximation (AMA), integrating Eq.~(\ref{eq:KSconteq}) over atom $i$, we are left with a simple equation for the time evolution of the atomic spins,
\begin{equation}
\frac{\partial \mathbf{S}_i}{\partial t}(t)=-\gamma \mathbf{S}_i(t) \times \mathbf{B}_{i},
\label{eq:slowvar}
\end{equation}
where $i$ denotes atomic index and $\mathbf{B}_i$ the effective field which the atomic spin, $\mathbf{S}_i$, experiences.

\subsection{Parametrization}
An accurate approach for performing spin dynamics and for calculating effective fields acting on the atomic spins, is to perform a constrained DFT calculation at each time step using local constraining fields. This has been done for systems consisting of a few atoms by {\'U}jfalussy \textit{et al.}\cite{Ujfalussy2004} where spin dynamics of a finite Co chain along a Pt(111) surface step edge was simulated. While accurate, the approach is computationally fairly cumbersome and much can be gained by working with a parametrization of the KS Hamiltonian. Such an approach was suggested by F$\textrm{\"{a}}$hnle \textit{et al.}\cite{Fahnle2005} where a gradual trade off between accuracy and computational requirements is possible. By using a spin-cluster expansion method, the effective field including exchange, magnetocrystalline anisotropy, dipolar and external field contributions were parametrized. By increasing the number of parameters in the parameterization, accuracy was increased toward \emph{ab initio} accuracy at the same time as the computational requirements increased.

We adopt a similar approach, in the sense that the energy of the system is parametrized and the dynamics is simulated for the parametrized Hamiltonian. Here, we present a general parametrization in terms of the atomic moments, $\mathbf{m}_i$, instead of the atomic spins, $\mathbf{S}_i$. For 3d-systems, the atomic moment is dominated by the spin moment contribution and the effect of spin-orbit coupling is small. For systems such as actinides, orbital moments are larger and the total atomic moment must be considered in dynamical simulations of the magnetization. This would involve replacing $\mathbf{S}_i$ for $\mathbf{J}_i$ in many of the expressions presented in this paper, but for simplicity we have not done so, since most of the examples presented here are for magnetism among 3d elements, where the orbital moment can be neglected.
In the KS Hamiltonian, we neglected dipolar interactions and spin-orbit coupling. Dipolar interactions are small and included separately in the parametrized Hamiltonian. Spin-orbit coupling gives rise to a magnetocrystalline anisotropy which also is included separately in the generalized Hamiltonian. The effective field, $\mathbf{B}_i$, on each atom is calculated from 
\begin{equation}
\mathbf{B}_i=-\frac{\partial \mathscr{H}}{\partial \mathbf{m}_i}.
\label{eq:heisenberg1}
\end{equation}
The parametrized Hamiltonian is composed of the following terms,
\begin{equation}
\mathscr{H}=\mathscr{H}_{\textrm{iex}}+\mathscr{H}_{\textrm{ma}}+\mathscr{H}_{\textrm{dd}}+\mathscr{H}_{\textrm{ext}},
\label{eq:heisenberg2}
\end{equation}
where for the first term, which represents interatomic exchange interactions, we use the classical Heisenberg Hamiltonian,
\begin{equation}
\mathscr{H}_{\textrm{iex}}=-\frac{1}{2}\sum_{i\neq j}J_{ij}\mathbf{m}_i\cdot\mathbf{m}_j,
\end{equation}
where $i$ and $j$ are atomic indices, $\mathbf{m_i}$ the classical atomic moment and $J_{ij}$ the strength of the exchange interaction. The second term in Eq.~(\ref{eq:heisenberg2}) represents the magnetocrystalline anisotropy and can take several forms. For a uniaxial anisotropy we have the dominant contribution of the form,
\begin{equation}
\mathscr{H}_{\textrm{ma}}=K\sum_{i}(\mathbf{m}_i\cdot \mathbf{e}_K)^2,
\label{eq:aniso}
\end{equation}
where $\mathbf{e}_K$ is the direction of the anisotropy axis and $K$ the strength of the anisotropy field.
The third term,
\begin{equation}
\mathscr{H}_{\textrm{dd}}=-\frac{1}{2}\sum_{i\neq j}Q_{ij}^{\mu \nu}m_i^{\mu}m_j^{\nu},
\end{equation}
represents dipolar interactions. Here $\mu$ and $\nu$ are coordinate indices and $Q_{ij}^{\mu \nu}$ is given by,
\begin{equation}
Q_{ij}^{\mu \nu}=\frac{\mu_0}{4\pi}(3R_{ij}^\mu R_{ij}^\nu-\delta_{\mu \nu}R_{ij}^2)R_{ij}^{-5},
\end{equation}
where $R_{ij}$ is the distance between atomic moments $i$ and $j$.
Dipolar interactions are long range and important for the long wave length excitations. The interaction can be neglected in studies of short wave length excitations. For finite systems dipolar interactions lead to a shape anisotropy. For a thin film the shape anisotropy can be modeled by a term similar to Eq.~(\ref{eq:aniso}),
\begin{equation}
\mathscr{H}_{\textrm{shape}}=K_{\textrm{shape}}(\mathbf{\bar{m}}\cdot \mathbf{e}_{\textrm{shape}})^2,
\label{eq:shapeaniso}
\end{equation}
where $\mathbf{e}_{\textrm{shape}}$ is the out-of-plane direction of the film, $\mathbf{\bar{m}}$ is the average magnetic moment of the system and $K_{\textrm{shape}}$ is the strength of the shape anisotropy. The last term of Eq.~(\ref{eq:heisenberg2}),
\begin{equation}
\mathscr{H}_{\textrm{ext}}=-\mathbf{B}_{\textrm{ext}}\cdot \sum_{i}\mathbf{m}_i,
\end{equation}
is the Zeeman term and describes the interaction of the magnetic system with an external magnetic field.

In our approach we use the parametrized Hamiltonian, Eq.~(\ref{eq:heisenberg2}), combined with Eq.~(\ref{eq:slowvar}) which describes the time evolution of the magnetization for a system which is dominated by the spin moment. Parameters for the parametrized Hamiltonian are obtained by a mapping from a DFT ground state calculation. 
The most widely used approach is through the Liechtenstein-Katsnelson-Gubanov method (LKGM)\cite{Liechtenstein87} which is based on the magnetic force theorem where parameters are obtained from small angle perturbations from the ground state. At low temperatures, where the inter-atomic angles between the atomic spins are small, the parameters can be considered accurate. For the paramagnetic state one may \newtext{instead} extract Heisenberg exchange parameters by means of the generalized perturbation method (GPM)\cite{Ruban2004} for a disordered local-moment (DLM) state treated within the coherent potential approximation (CPA). This method provides a more accurate description of the high temperature region.

\subsection{Damping}
When the atomic spins evolve from the dynamics of Eq.~(\ref{eq:slowvar}), energy and angular momentum dissipates via a range of mechanisms. The different mechanisms which lie behind this damping have e.g. been studied in Refs. \onlinecite{Elliot1954,Yafet1963,Kambersky1970,Kunes2002,Fahnle2005b,Fahnle2006,Kambersky1976,Stiles2007,Ho2004,Beaurepaire2004}.
The effect of the different damping mechanisms is normally included by adding a phenomenological term to Eq.~(\ref{eq:slowvar}), which yields the LLG form, \comment{parentheses}
\begin{equation}
\frac{\partial \mathbf{S}_i}{\partial t}=-\gamma \mathbf{S}_i \times \mathbf{B}_i +\frac{\alpha}{S_i}  \mathbf{S}_i \times \frac{\partial\mathbf{S}_i}{\partial t} \,,
\label{eq:LLGdamping}
\end{equation}
where $\alpha$ is the the damping coefficient and $S_i$ is the size of the spin $\mathbf{S}_i$. 
For numerical reasons we use the Landau-Lifshitz (LL) \newtext{form of} damping term, and hence Eq.~(\ref{eq:LLGdamping}) is replaced by \comment{parentheses}
\begin{equation}
\frac{\partial \mathbf{S}_i}{\partial t}=-\gamma\mathbf{S}_i \times \mathbf{B}_i-\gamma\frac{\alpha}{S_i} (\mathbf{S}_i \times (\mathbf{S}_i \times \mathbf{B}_i)).
\label{eq:LLdamping}
\end{equation}

\section{Finite temperature modeling}

\begin{figure}
\begin{center}
\includegraphics*[width=0.45\textwidth]{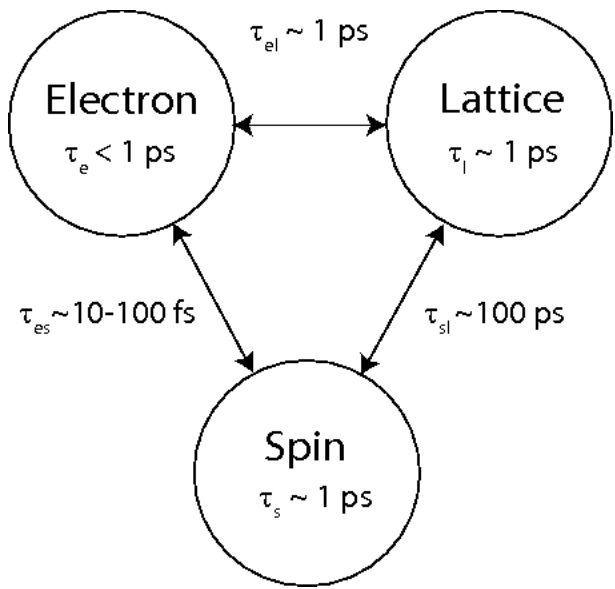}
\caption{The dynamic behavior of a magnetic solid can be understood in terms of three thermodynamic reservoirs and interactions that exchange energy between the reservoirs. In the figure we included approximate relaxation times within the reservoirs and between the reservoirs.}
\label{fig:thermores1}
\end{center}
\end{figure}

Most of the systems we are interested in simulating with the here presented method can conceptually be understood in terms of three thermodynamic subsystems; the spin system, the electronic system and the lattice (Fig.~\ref{fig:thermores1}). The different reservoirs can be identified in measurements of specific heat. Each of these subsystems can be seen as reservoirs for energy and angular momentum.

A division of the magnetic solid as such, into three thermodynamic reservoirs, is not free from complications, especially the division of the electronic system and the spin system which both are manifestations of the nature of electrons. \label{sec:thermo}
It is important to note that the elementary excitations of the spin system carry an angular momentum of $\hbar$. Any transfer of energy to or from the spin system must be accompanied by a transfer of angular momentum. The necessity of angular momentum conservation is often a bottleneck of the transfer of energy between the subsystems.

The processes that carry energy and angular momentum between the subsystems are defined by the way the division is made. The total Hamiltonian for the magnetic solid carries terms that mix the subsystems and these are the processes which are responsible for the energy and angular momentum exchange between the subsystems.
Relaxation rates between the reservoirs are associated with the characteristic energies of the interactions that mediate the coupling between the reservoirs.\cite{Stohr2} These time scales have been measured in experiments. The electron-lattice relaxation time, $\tau_{\textrm{el}}$, is of the order of picoseconds (ps). The spin-lattice relaxation time, $\tau_{\textrm{sl}}$, is of the order of 100~ps and the spin-electron relaxation time, $\tau_{\textrm{es}}$, has been found in recent pump-probe experiments to be of the order of 100~fs.\cite{Rasing,Stohr} Relaxation within the spin system ($\tau_{\textrm{s}}$) and within the lattice ($\tau_{\textrm{l}}$) are expected to take place on times scales of the order of picoseconds whereas the electron-electron relaxation ($\tau_{\textrm{e}}$) takes place on a subpicosecond time scale.

In order to describe atomistic spin dynamics at finite temperatures in simulations, the spin system must be coupled to a thermal reservoir in such a way that energy may be transferred into and out of the system. We will start by showing how a single thermal reservoir can be coupled to the spin system and later generalize the discussion to several thermal reservoirs. \label{sec:temperature}

\subsection{One thermal reservoir}
For a discussion on stochastic and deterministic methods of including temperature, see Ref.~\onlinecite{Antropov1996}. One way of introducing a coupling to a thermal reservoir, which is adopted here, is through Langevin Dynamics (LD), which is standard in finite temperature micromagnetic simulations.\cite{GarciaPalacios1998, Novotny1, Novotny2, Novotny3} In our approach, excitations are generated by performing classical rotations of single atomic spins in such a way that the energies of the atomic spins satisfy Boltzmann statistics. As a practical method, either Monte Carlo (MC) or LD methods may be used for obtaining a finite temperature equilibrium configuration.

Thermal excitations are generated by adding a stochastic field, $\mathbf{b}_i$, to the effective field, $\mathbf{B}_{i}$, on each atom, $i$. The random field is assumed to be a Gaussian stochastic process with the following statistical properties,
\begin{equation}
\langle b_{i,\mu}(t)\rangle=0, \quad \langle b_{i,\mu}(t)b_{j,\nu}(s)\rangle =2D\delta_{\mu \nu}\delta_{ij}\delta(t-s),
\label{eq:bfield}
\end{equation}
where $\mu$ and $\nu$ are the Cartesian coordinates of the field and where $D$ is the strength of the thermal fluctuations. 
The Kronecker deltas in Eq.~(\ref{eq:bfield}) state that the different Cartesian components of $\mathbf{b}_{i}$ are unrelated and that the random fields acting on different magnetic moments $i$ are independent. The Dirac delta states that the autocorrelation time of $\mathbf{b}_{i}$ is much smaller than the rotational response of the system.

As a technical note we mention that we have chosen to add the stochastic field to the effective field in both the precessional term and the damping term, resulting in the following equation:
\begin{equation}
\frac{\partial \mathbf{S}_i}{\partial t}=-\gamma (\mathbf{S}_i \times (\mathbf{B}_{i}+\mathbf{b}_{i}(t)))-\gamma \frac{\alpha}{S_i} (\mathbf{S}_i \times (\mathbf{S}_i \times (\mathbf{B}_{i}+\mathbf{b}_{i}(t)))).
\label{eq:sllg}
\end{equation}
Eq.~(\ref{eq:sllg}) is a stochastic differential equation (SDE) as opposed to regular ordinary differential equations (ODE) and require an interpretation rule.\cite{Kloeden} In Appendix \ref{sec:appb} we present a derivation of the amplitude of the stochastic field, $D$, required to achieve thermodynamic consistency.

\begin{figure}
\includegraphics*[width=0.45\textwidth]{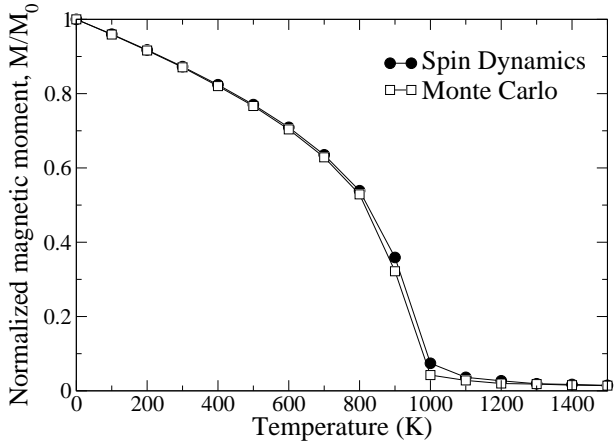}
\caption{Comparison of equilibrium magnetization versus temperature for a periodic $20\times 20\times 20$ bcc Fe system for SD and MC. }
\label{fig:sdmc}
\end{figure}

\begin{figure}
\includegraphics*[width=0.45\textwidth]{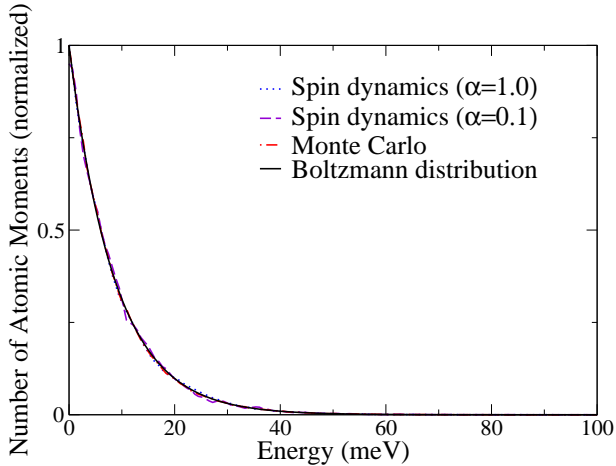}
\caption{ (Color online) Histogram of the energy distribution of the atomic spins of a $20\times 20\times 20$ bcc Fe system within MC and SD simulations at 100K. The SD simulation is is done for two different damping parameters and data is obtained at equilibrium.}
\label{fig:Hist1b}
\end{figure}

At equilibrium, MC and Spin Dynamics (SD) give identical results for a number of properties. MC can actually be used as a way of benchmarking SD simulations. In Fig.~\ref{fig:sdmc} we plot the saturation magnetization for MC and SD for bcc Fe versus temperature. Simulations are performed on a $20\times 20\times 20$ bcc system using four coordination shells in the Heisenberg term. The Heisenberg exchange parameters were calculated from first-principles theory and MC and SD are seen to give identical results. In Fig.~\ref{fig:Hist1b} we plot the energy distribution of the moments for MC simulations and SD simulations with two different damping parameters. These distributions coincide perfectly with the Boltzmann distribution. 

\begin{figure}
\includegraphics*[width=0.45\textwidth]{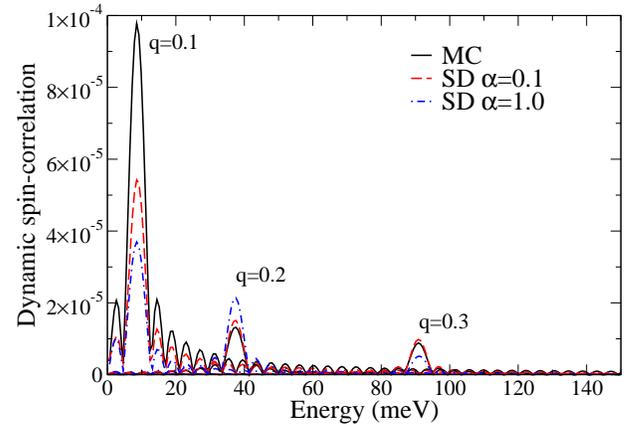}
\caption{(Color online) The first three excitation peaks in $S(q,\omega)$ for a periodic, $20\times 20\times 20$ bcc, Fe system at 100K. The size of the peaks in the SD simulation vary depending on the damping parameter.}
\label{fig:LE}
\end{figure}

\begin{figure}
\includegraphics*[width=0.45\textwidth]{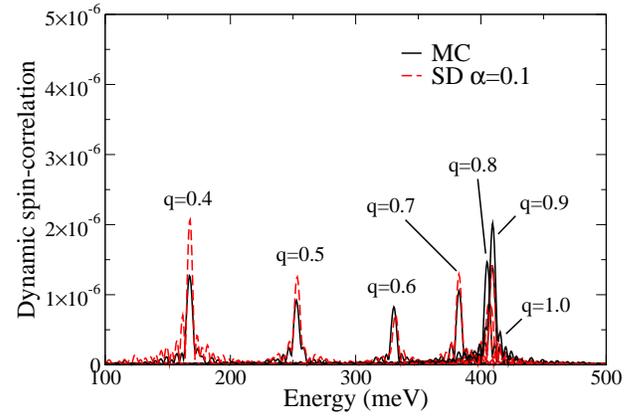}
\caption{(Color online) Same as in Fig.~\ref{fig:LE} but in the high energy region. Comparison of MC and SD with $\alpha=0.1$.}
\label{fig:HE1}
\end{figure}

\begin{figure}
\includegraphics*[width=0.45\textwidth]{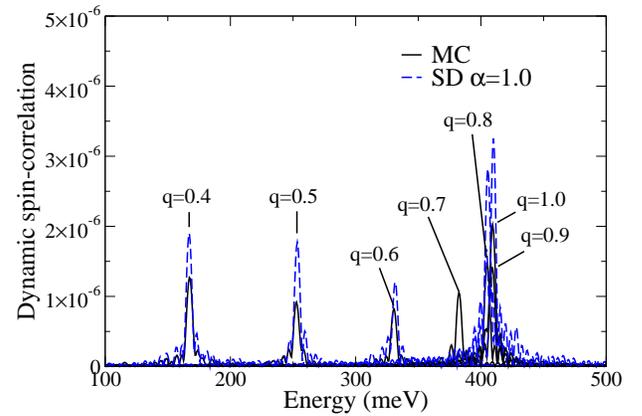}
\caption{(Color online) Same as in Fig.~\ref{fig:LE} but in the high energy region. Comparison of MC and SD with $\alpha=1.0$.}
\label{fig:HE2}
\end{figure}

Differences between SD and MC are more subtle and appear first in comparisons of spin-correlation. Fig.~\ref{fig:LE}-\ref{fig:HE2} illustrate the dynamic spin-correlation function $S(\mathbf{q},\omega)$, which is described below in Section~\ref{sec:sqw}, calculated for equilibrium states generated by MC and SD simulations with different damping parameters. For Fe realistic damping parameters are of the order 0.005-0.1.\cite{Stamm2005} The comparisons shows that the MC generated equilibrium has a larger amount of high energy/large momentum excitations and a lower amount of low energy excitations than the SD system with damping $\alpha=0.01$. By increasing the damping parameter in SD the excitation content is modified and for large enough damping parameter the number of high energy excitations exceeds that found for the MC equilibrium. 


\subsection{Several thermal reservoirs}
We have now described how the magnetic system can be connected to one thermal reservoir. In order to properly represent the spin dynamics of a system, where the system can be decoupled into three thermal reservoirs as described in Section~\ref{sec:thermo}, we present a method of connecting the spin system to several thermodynamic reservoirs. The relaxation time between the electronic system and the lattice is of the order of picoseconds. Hence, a distinction between the electronic reservoir and the lattice is only necessary when studying dynamics with resolution higher than picoseconds. To describe the interaction between the electronic system, the lattice and the spin system, it is natural to propose a two-damping model. The intent is to capture the interaction between the spin system and the lattice with one damping parameter and to capture the interaction between the spin system and the electrons with a second damping parameter. A third parameter is also needed and describes the transfer of energy between the electrons and the lattice. Until there is more knowledge on how these parameters can be calculated, we use parameters obtained by fitting to pump probe experiments which display all these processes.

\begin{figure}
\includegraphics*[width=0.45\textwidth]{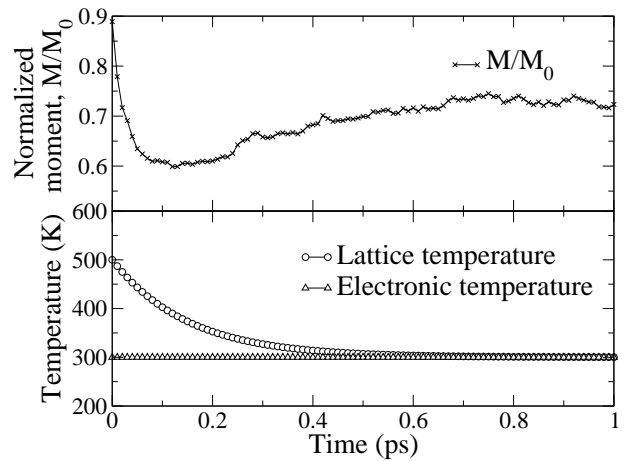}
\caption{Simulation of a pump probe experiment. Simulations were done on a $10\times 10\times 10$ bcc Fe system. The lower panel shows the assumed electron and lattice temperatures. The lattice temperature is constant at 300~K. The electronic temperature decays from an initial 500~K to 300~K with a decay time of 150~fs. The upper panel shows the normalized spin moment of the system.}
\label{fig:koop}
\end{figure}

We thus proceed by introducing two Gilbert damping terms which are added to the equation of motion for the atomic moments,  \comment{parentheses}
\begin{eqnarray}
\frac{\partial \mathbf{S}_i}{\partial t}&=&-\gamma \mathbf{S}_i \times (\mathbf{B}_i+\mathbf{b}_i(t))- \nonumber \\
&&-\gamma \frac{\alpha_e}{S_i} \mathbf{S}_i \times (\mathbf{S}_i \times (\mathbf{B}_i+\mathbf{b}_i(t)))- \nonumber \\
&&-\gamma \frac{\alpha_l}{S_i} \mathbf{S}_i \times (\mathbf{S}_i \times (\mathbf{B}_i+\mathbf{b}_i(t)))
\label{eq:main}
\end{eqnarray}
where $\alpha_{\textrm{e}}$ and $\alpha_{\textrm{l}}$ are the damping parameters which correspond to an energy transfer from the spin system to the electrons and to the lattice, respectively. Eq.~(\ref{eq:main}) describes how energy dissipates from the system through two channels. The temperature of the reservoirs are given by $T_{\textrm{e}}$ and $T_{\textrm{l}}$. In equilibrium the temperature of all three thermodynamic reservoirs are the same, i.e. $T_{\textrm{e}}=T_{\textrm{l}}=T_{\textrm{s}}$. In our treatment the amplitude of the thermal fluctuations $\mathbf{b}_i$ is given by, 
\begin{equation}
D = D_{\textrm{e}}+D_{\textrm{l}}, 
\label{eq:d}
\end{equation}
where
\begin{equation}
D_{x} = \frac{1}{1+(\alpha_{\textrm{e}}+\alpha_{\textrm{l}})^2} \frac{k_{\textrm{B}} T_{x}}{\gamma m_s} \alpha_{x}, 
\label{eq:dllg2T}
\end{equation}
where $x=\textrm{e,l}$ corresponds to electron or lattice effects. These amplitudes correspond to the equilibrium thermal fluctuations. Assuming a constant flux of energy from the reservoirs to the spin system, we use these amplitudes in our dynamic simulations. For practical simulations assumptions need to be made on the initial temperatures, $T_{\textrm{e}}$ and $T_{\textrm{l}}$, and the relaxation between the electrons and the lattice. Typical pump-probe experiments as reported in Ref.~\onlinecite{Koopmans2005b} can be simulated by assuming that the lattice is an infinitely large thermal reservoir with constant temperature $T_{\textrm{l}}$. Further, we assume that the electron reservoir is a thermal reservoir much smaller than the lattice, but much larger than the spin system with a temperature that evolves with time as $T_{\textrm{e}}(t)=T_{\textrm{l}}+T_{\textrm{e,init}}\cdot \textrm{exp}(-t/\tau_{\textrm{el}})$ and where $T_{\textrm{e,init}}$ is the initial temperature. As an application of our two-damping model we address recent pump probe experiments\cite{Koopmans2005b}, where the magnetization dynamics following optical excitation of a Ni film has been interpreted in terms of the three thermal bath model. We consider a test system of bcc Fe with four coordination shells, as described above, and are able to reproduce the trends found in experiments. Our simulated magnetization is shown in Fig.~\ref{fig:koop}. In the top graph it is seen that the magnetization initially decreases and after some time ($\sim$ 0.5~ps) it stabilizes at a value $\sim$ 70 \% of the initial value. This behavior is in qualitative agreement with the measured data in Ref.~\onlinecite{Koopmans2005b}.

\section{Extracting information}
A challenge in practical simulations of magnetization dynamics is extracting, visualizing and comprehending results. A simulation of the time development of the magnetic moment, $\mathbf{m}_i$, of $N$ atoms over $M$ time steps, generates data on the form $m_i^j(t_k)$ where $j=x,y,z$, $i=[1,N]$ and $t_k=[1,t_M]$. For a typical simulation this amounts to an unmanageable amount of data which is difficult to store. In order to analyze and comprehend the meaning of the data it must be compressed into variables that capture the state and evolution of the system. By doing this on the fly during the simulation, computational time and storage requirements are greatly saved. \label{sec:extracting}Below we analyze in this way trajectories of the atomic spins, average moment, spin-correlations and the energy distributions in simulations of spin-dynamics.

\subsection{Trajectories}
\begin{figure}
\includegraphics*[width=0.45\textwidth]{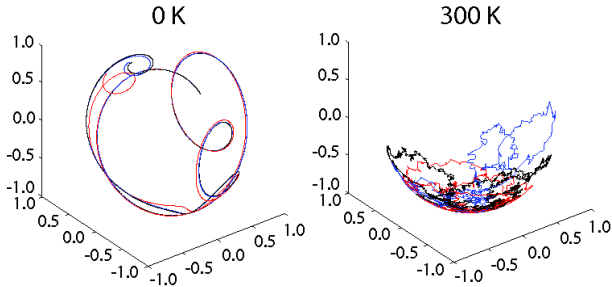}
\caption{(Color online) The trajectory of an atomic spin for a duration of 100 fs is shown. Simulations are performed on a $10\times 10\times 10$ bcc Fe system. On the left hand side the trajectory is shown for one atomic spin in a non-equilibrium system at 0K where the atomic spin directions are completely randomly distributed. On the right hand side we show the trajectory of an atomic spin at a 300~K equilibrium. Simulations are performed for different sizes of times steps in the numerical method: $10^{-18}$ (black), $10^{-17}$ (blue) and $10^{-16}$ (red). 
}
\label{fig:traj}
\end{figure}

In Fig.~\ref{fig:traj} we show the trajectories of individual atomic moments. The simulations are performed on a $10\times 10\times 10$ system of bcc Fe with periodic boundary conditions. The duration of all simulations are 100~fs and the damping is $\alpha=0.1$. On the left hand side we present a simulation at 0~K where the initial spin distribution is random. Trajectories are presented for three different step sizes in the numerical scheme where Heun's scheme was used. With this scheme and for this particular simulation, step sizes as small as 1-10 attoseconds are required to produce accurate trajectories on a time scale of 100~fs. On the right hand side we present the trajectory of one atomic spin of a system in a 300~K equilibrium. We present simulations for three different step sizes. At finite temperatures individual trajectories do not carry much information because of thermal fluctuations. By viewing snap shots or sequences of snap shots of the spin configuration over the entire system or parts of the system, valuable information on correlations and domain formation can be visualized.\cite{fysik4}  

\subsection{Average magnetic moment}
Averages are fundamental quantities of a magnetic system. With the data from a spin dynamic simulation averaging can be performed over space, time, different random number sequences in the Langevin equations or over different initial states. Thermal (ensemble) averages are often desired and can be calculated in different ways depending on the system and process.

\begin{figure}
\includegraphics*[width=0.45\textwidth]{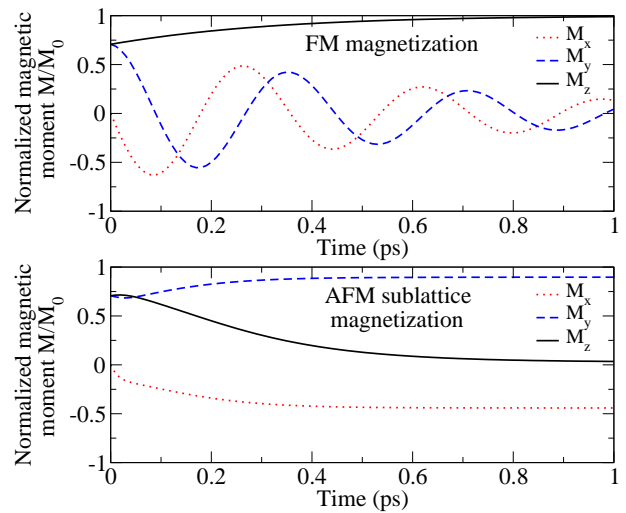}
\caption{(Color online) Relaxation in an easy-axis anisotropy. The upper panel shows the relaxation of a ferromagnet in an easy-axis uniaxial anisotropy and the second panel shows the relaxation of an anti-ferromagnet in an easy-axis the same anisotropy. For the ferromagnet we plot the evolution of the average magnetization and for the anti-ferromagnet we plot the evolution of the average magnetization of one sublattice.}
\label{fig:AFM}
\end{figure}

Space averaging over all atoms in the system gives the average magnetization. If all atoms are equivalent, such an average may be taken as a thermal average. Space averaging may also be performed separately over different sublattices or separately over sets of equivalent atoms in order to understand the the dynamical behavior of certain parts of a system. This is useful when studying antiferromagnets (AFM) or alloys. In Fig.~\ref{fig:AFM} we show the relaxation of a ferromagnet and an antiferromagnet in an uniaxial easy-axis anisotropy, following a sudden 45 degree change with respect to the anisotropy axis. For the ferromagnet (upper panel) we show the evolution of the average magnetization whereas for the antiferromagnet (lower panel) we show the evolution of the average magnetization of a sublattice. In order to understand the switching behavior of an anti-ferromagnet the behavior of each sublattice and their mutual interaction plays an important role. For future large scale simulations where the spatial variation of the magnetization over larger length scales is interesting, space averaging may be performed over several limited spatial regions of the system, producing a more coarse grained picture of the spin dynamics of the system. Fig.~\ref{fig:AFM} also shows that the switching of the AFM is faster than for the FM.

Time averaging is useful for smearing out random fluctuations. In equilibrium, time averaging may be performed over unlimited time. When studying dynamic processes, time averaging must be performed over sufficiently short time intervals in comparison to the time scale of the dynamic process. For systems such as spin glasses, where each atomic moment is unique, thermal averaging may be done by performing an averaging over time. In particular for spin glasses, which often are out of equilibrium, such a time averaging must be performed over sufficiently short time intervals. 
For systems with bond- or site-disorder, averaging can be done over different configurations of exchange parameters respectively magnetic atoms in the lattice. Dilute magnetic semiconductors are a manifestation of site-disordered systems. Among different classes of spin glasses there are systems possessing either bond- or site-disorder.
Often a combination of space and time averaging is useful. In space averaging the number of averaging terms is limited by the finite size of the system. Finite size effects themselves have effects on the system which are interesting to study. For small systems space averaging may become insufficient and can be compensated by time averaging.

Another type of averaging is averaging over identical simulations but with different random number sequences in the Langevin equations. For equilibrium simulations this is similar to time averaging. This type of averaging is however very time consuming since the same simulation must be repeated several times. The technique is best used in combination with space and time averaging.

For some specific simulations one might also consider sampling over different initial configurations. The different but equivalent initial configurations could be generated with Monte Carlo or with Spin Dynamics using different initial configurations.

\subsection{Correlations between magnetic moments}
In addition to the trajectories or the absolute directions of atomic moments, correlations or relative directions between atomic moments, provide fundamental information on the system (see Refs.~\onlinecite{Chen1994, Tsai2000, Tao2005}). The correlation function can be defined as
\begin{equation}
C^k(\mathbf{r}-\mathbf{r'},t)=\langle m^k_{\mathbf{r}}(t)m^k_{\mathbf{r'}}(0)\rangle - \langle m^k_{\mathbf{r}}(t)\rangle \langle m^k_{\mathbf{r'}}(0)\rangle,
\end{equation}
where $\langle ... \rangle$ denotes ensemble averages and may be performed according to the previous section. The first term on the right hand side is the overlap and contains information on the magnetic order of the system. \label{sec:sqw}

In order to evaluate the spin wave excitation content of a system one may calculate $S(\mathbf{q},\omega)$ by performing a space and time fourier transform of the spin-spin correlation,
\begin{equation}
S^{k}(\mathbf{q},\omega ) = \frac{1}{N\sqrt{2\pi}}\sum_{\mathbf{r},\mathbf{r'}}e^{i\mathbf{q}\cdot (\mathbf{r}-\mathbf{r'})}\int_{-\infty}^{+\infty}e^{i\omega t}C^k(\mathbf{r}-\mathbf{r'},t) dt,
\end{equation}
where $N$ is the number of terms in the summation. Fig.~\ref{fig:LE}-\ref{fig:HE2} show $S(\mathbf{q},\omega)$ for an equilibrium system. For dynamical processes it is interesting to study how the spin wave content changes with time. Such a calculation was performed in Section~\ref{sec:switching} on bcc Fe in an ultra strong switching field. For this process the time scale of the switching process was too fast to allow for an accurate $S(\mathbf{q},\omega)$ calculation at different points in time of the dynamic process. The calculation was instead performed by taking snapshots of the configuration of the system at different points in time during the dynamic process. Each snap shot then serves as an initial state in a zero damping simulation where the dynamic spin-correlation is calculated. This procedure works as long as the system does not exhibit any strong spin-wave instabilities, which may change the spin-wave content at zero damping.


\subsection{Energy distributions}
In equilibrium the energy distribution of the spin-moments follows a Boltzmann distribution, as shown above. The energy of spin $i$ is given by
\begin{equation}
E_i=-\mathbf{m}_i\cdot\mathbf{B}_i+|\mathbf{m}_i||\mathbf{B}_i|.
\end{equation}
In this expression parallel coupling between moment and local effective field is set to zero energy. Fig.~\ref{fig:Hist1b} shows a comparison between MC and SD for different values of the damping parameter. During dynamical processes the distribution changes. In analog with the spin wave content one may calculate the change in the energy distribution at different points in times during a dynamic process.

\subsection{Direct visualization}
\begin{figure}
\includegraphics*[width=0.40\textwidth]{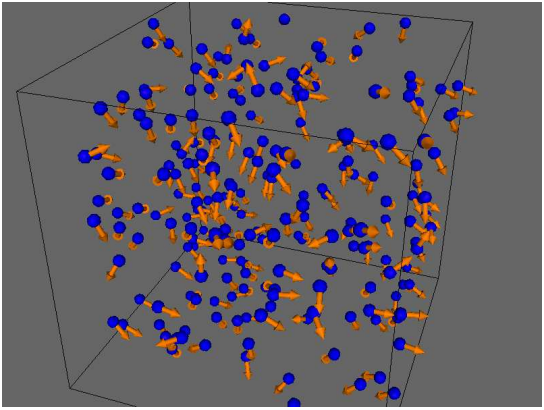}
\caption{(Color online) Snap-shot of a simulation of Mn doped GaAs (5\% concentration). Only Mn atoms are shown, in blued color. The orange arrows indicate the size and direction of the Mn atomic spins.}
\label{fig:GaMnAs}
\end{figure}

Perhaps the most natural illustration of a spin-dynamics simulation is to present a real time visualization of how the spins relax during the simulation. In Fig.~\ref{fig:GaMnAs} we present an example of this where a snap-shot of the spin-configuration of Mn doped GaAs, a diluted magnetic semiconductor, is shown. Only the magnetic Mn atoms are shown, the nonmagnetic Ga- and As-atoms are not visualized. The data is from a simulation where the Mn concentration was 5\%.  The temperature of the simulation is $T=100$~K, which is below the ferromagnetic ordering temperature. The snap-shot illustrates the extent of correlation on short distances, whereas the value of the global magnetization is better obtained as a thermal average.

\section{Applications}
An atomistic approach to spin dynamics is necessary for various classes of problems. One class of problems are systems at extreme conditions such as extreme external magnetic fields, where high energy short wavelength magnons are excited. This case will be treated in the following section. Another class of problems are systems with complex magnetic ordering on an atomic scale which cannot be treated properly within micromagnetism such as dilute magnetic systems,\cite{Hellsvik2007DMS,Skubic2007SG} anti-ferromagnets and spin-spirals. A third class of problems concern systems with complex chemical ordering or nano-structured materials.\cite{Skubic2007AF, Skubic2007bias} In the following sections we present simulations of bcc Fe in large magnetic switching fields. We also illustrate different techniques for visualizing a dynamic magnetization process. \label{sec:applications}

\section{Magnetic Switching}
LLG theory relies on a macrospin approximation where the magnitude of the macrospin is assumed constant. It has previously been shown that for large anisotropies, the macrospin picture breaks down due to the appearance of spin-wave instabilities which alter the size of the macrospin. Cases where the macrospin approach breaks down may present interesting areas of application of atomistic spin dynamics. In this section we address magnetic switching in an external field. We show that the macrospin approximation remains valid up to very high switching field strengths. However, at extremely high switching fields, over 100~T, the approximation finally breaks down. Here, we use this limit to illustrate the use of the atomistic spin dynamics method. By using bulk Fe as a model system we describe the switching process from atomistic considerations. Furthermore we determine the size of the external field when LLG theory breaks down. \label{sec:switching}

Magnetic switching is the process of moving a system from one stable magnetic configuration to another and is fundamental for any system where a magnetic state is used for storing and retrieving information. The switching process involves an excitation of the system followed by a relaxation into a new stable configuration. In this section we address magnetic switching induced by an external \tobedeleted{switching} field\tobedeleted{. By applying an external field}, \newtext{in presence of which} energy is \tobedeleted{deposited directly} \newtext{transferred}  into the magnetic system through the Zeeman term. After this excitation, the system relaxes into a new stable configuration.

Recent field pulse magnetic switching experiments explore magnetic switching in field pulses of the order of 35~T,\cite{Stohr2} a factor of 1000 higher than field pulses in conventional switching experiments. As an external field is applied to the system, the spin-wave spectrum of the system is shifted by a Zeeman term which is either positive or negative depending on if the external field is applied parallel or antiparallel with the magnetization. Hence, the external field yields an excitation directly into the spin system. The system is then driven to equilibrium with the other thermal reservoirs by different damping processes. Both the orientation and magnitude of the macro-spin changes in a process which will be the focus of this section. For weak switching fields, the change in magnitude of the macrospin is negligible and the results approach the LLG theory. For large fields (100 T) comparable to the exchange field (1000 T), the change in magnitude is significant and an atomistic approach is necessary for describing the process.

At zero temperature our atomistic model coincides with the LLG picture since the system lacks high energy thermal excitations. At finite temperatures the system contains these high energy thermal excitations which alter details of the switching model. Let us now venture to an atomistic picture of the field induced switching process at finite temperatures. Since the main purpose is a qualitative description of the switching process, the atomistic simulations are again performed on bcc Fe using four coordination shells in the Heisenberg Hamiltonian. Simulations are performed on a system of $20\times 20\times 20$ bcc cells.

\begin{figure}
\includegraphics*[width=0.45\textwidth]{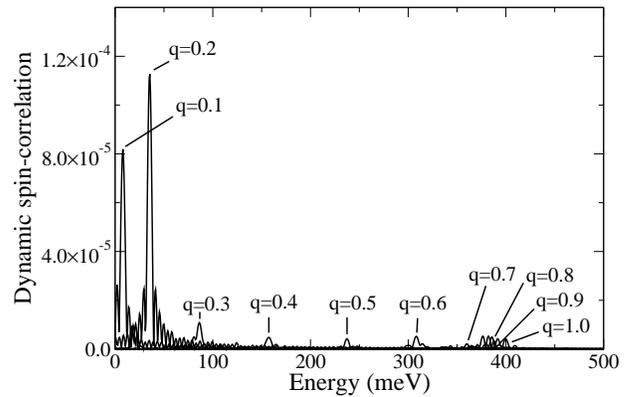}
\caption{Excitation content of bcc Fe in a 300~K equilibrium}
\label{fig:sqw}
\end{figure}
\begin{figure}
\includegraphics*[width=0.45\textwidth]{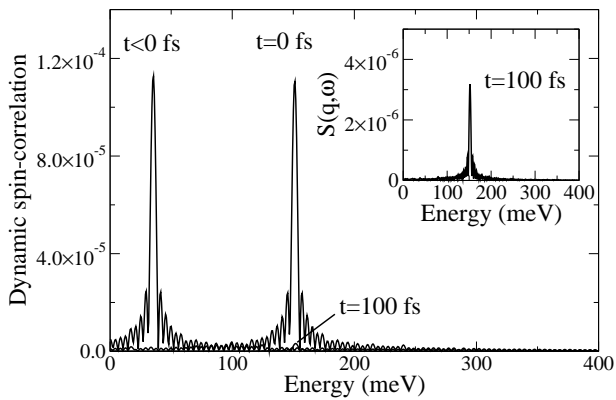}
\caption{The position and magnitude of the $\mathbf{q}=0.2\mathbf{e}_z (2\pi/a)$ peak as a 1000~T magnetic field is applied along the magnetization axis, i.e. z-axis, at $t=0$. The figure shows the original peak at $t<0$, the peak at $t=0$ when the external field is applied and the final peak (100~fs). The inset shows the final peak in blown up scale.}
\label{fig:singlesqw}
\end{figure}

Consider the system in equilibrium at 300~K. Fig.~\ref{fig:sqw} shows the calculated $S(\mathbf{q},\omega)$ of the equilibrium state, which gives an image of the excitations in the system. Now consider what happens when a constant magnetic field of 1000~T is applied. We use an exaggerated large switching field in order to demonstrate the effect. As a first case consider the external field being applied parallel to the magnetization. The external field shifts the positions of the excitations to higher energies. The process is illustrated in Fig.~\ref{fig:singlesqw} for the $\mathbf{q}=0.2\mathbf{e}_z (2\pi/a)$ excitation peak. The figure presents the original equilibrium peak position for $t<0$. At $t=0$ the 1000~T field is applied to the system resulting in a sudden shift of the peak. The external field adds to the exchange field which in turn increases the precessional torque on the atomic moments. This increases the frequency of precession and thereby the energy of the excitation. The system is now in a non-equilibrium state and relaxes with time through various dissipation processes to a new equilibrium. The damping torque brings the system to a new equilibrium with a larger saturation magnetization than the original equilibrium. One way to see this is that the external field increases the damping torque which in turn reduces the spread of the atomic moments and thereby increases the saturation magnetization. One can also regard the external field as a Zeeman contribution to the energy of the excitations. The damping term provides a path for energy and angular momentum to leave the system. Since the energy of the excitations was increased by the Zeeman contribution, excitations need to be removed from the system to restore the 300~K equilibrium. Fig.~\ref{fig:singlesqw} shows how the excitation peak shrinks significantly with time, leaving a barely visible peak, also  shown in the inset. We see that in an atomistic picture, the magnitude of the macrospin changes in the applied magnetic field. This change is neglected in LLG theory where the system remains unchanged as a magnetic field is applied parallel to the magnetization.
\begin{figure}
\includegraphics*[width=0.45\textwidth]{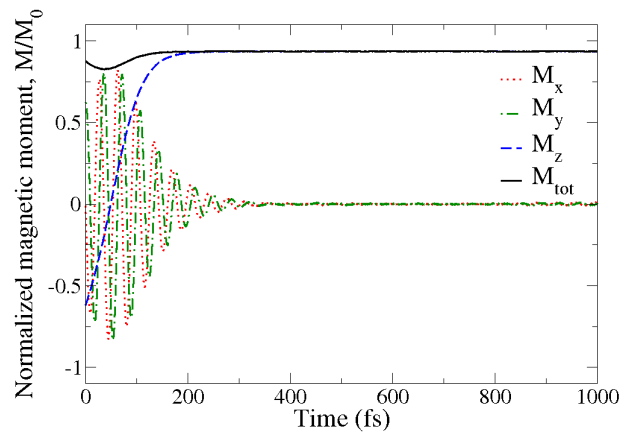}
\caption{(Color online) Magnetic switching process in a 1000~T constant external field applied 135 degrees with respect to the magnetization at 300~K. The Cartesian components and the magnitude of the normalized (0~K) magnetic moment are shown.}
\label{fig:T300H1000}
\end{figure}
\begin{figure}
\includegraphics*[width=0.45\textwidth]{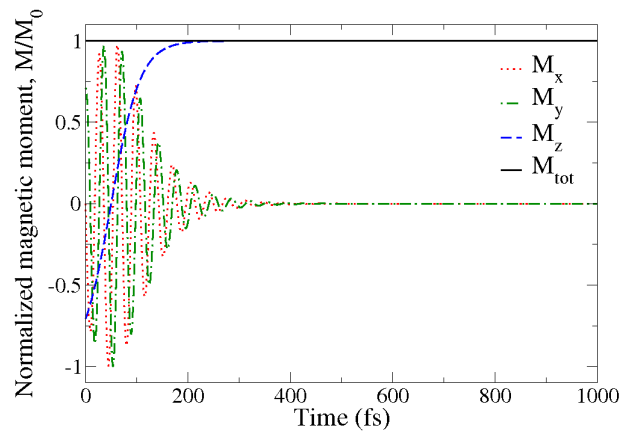}
\caption{(Color online) Magnetic switching process in a 1000~T constant external field applied 135 degrees with respect to the magnetization at 0~K. The Cartesian components and the magnitude of the normalized (0~K) magnetic moment are shown.}
\label{fig:T0H1000}
\end{figure}

\begin{figure}
\includegraphics*[width=0.45\textwidth]{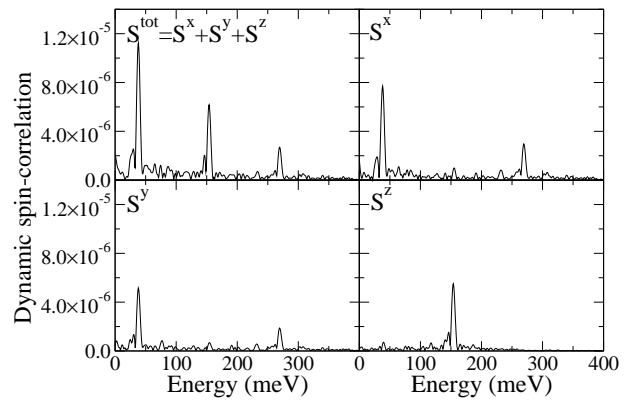}
\caption{The splitting of the $\mathbf{q}=0.5\mathbf{e}_z (2\pi /a)$ excitation peak as a 1000~T external field is applied at an angle of 135 degrees with respect to the magnetization.}
\label{fig:split}
\end{figure}

\begin{figure}
\includegraphics*[width=0.45\textwidth]{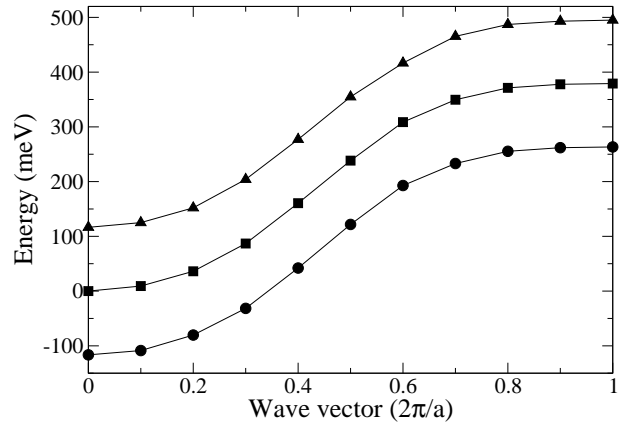}
\caption{Splitting of the excitation spectrum during the relaxation of the magnetic system in a 1000~T external field. The dispersion relationship for the three peaks in the upper left corner of Fig.14 are shown, where circles represent the low energy peak, triangles the high energy peak and squares the middle peak.}
\label{fig:splitfull}
\end{figure}

\begin{figure}
\includegraphics*[width=0.45\textwidth]{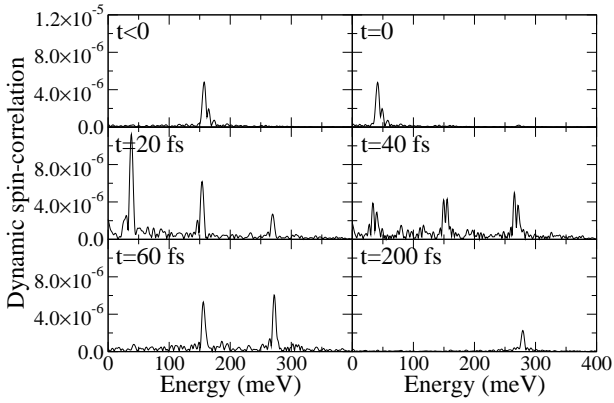}
\caption{Evolution of the $\mathbf{q}=0.5\mathbf{e}_z (2\pi /a)$ excitation peak during the switching process. At $t<0$, the peak is in its equilibrium position with its equilibrium magnitude. At t=0, a 1000~T field is applied at an angle of 135 degrees with respect to the magnetization. The peak is split in three. The negative shift is dominant in magnitude and the remaining two peaks are too small to be visible. Transfer of energy to the magnetic system leads to an increase in the peak size. As the switching proceeds the relative weight of the peaks change leading to a larger weight on the positively shifted peak. At this point, the energy transfer is reversed and energy is transferred out of the system, leading to a shrinking of all peaks. At the new equilibrium, the only peak remaining is the positively shifted peak. The magnitude of the peak is reduced compared to the original equilibrium peak.}
\label{fig:splitprocess}
\end{figure}

As a second case, let us now turn to a slightly more complex scenario - application of a 1000~T external field at an angle of 135 degrees with respect to the magnetization. This scenario eventually leads to a switching of the direction of the macrospin. It is interesting however to consider the path that the system takes to reach equilibrium in an atomistic picture. The $z$-axis is placed parallel with the external field and the initial magnetization lies along the $\frac{1}{\sqrt{2}}(\mathbf{e}_y-\mathbf{e}_z)$ direction. The switching processes for 300~K and 0~K are illustrated in Fig.~\ref{fig:T300H1000} and Fig.~\ref{fig:T0H1000}, respectively. For the 300~K case, there is a small reduction of the magnitude of the average magnetization during the switching process. We also see an oscillation of the $x$- and $y$-components of the macrospin, signaling the precession of the magnetization. Let us now analyze the switching process in more detail. As the field is applied to the system there is a splitting in the excitation spectrum. When the external field was applied parallel to the magnetization there was only a positive Zeeman contribution to the excitations. For an anti-parallel field we would find a negative Zeeman contribution whereas for a field applied in any other direction there is a negatively shifted peak, a positively shifted peak and a peak at the original position. This splitting is illustrated in Fig.~\ref{fig:split} for the $\mathbf{q}=0.5\mathbf{e}_z (2\pi /a)$ peak. In this plot the original peak is at $\sim 150$~meV. The magnitude of the peaks with respect to each other depends only on the angle of the magnetization with respect to the external field. Since the angle is 135 degrees in Fig.~\ref{fig:split}, it is the peak with the negative shift which dominates. The momentum dependence of the splitting of the excitation spectrum is illustrated in Fig.~\ref{fig:splitfull}. Note that this splitting is a feature of the non-equilibrium system. The central spectrum corresponds to the equilibrium excitation spectrum. Application of the 1000~T field has two main consequences. First, it excites the uniform motion of the magnetization or the average magnetization of the system. Secondly, within an atomistic model, it also splits the excitation peaks of the non-uniform magnons. Since the initial angle between magnetization and external field is 135 degrees, the lower branch dominates. This branch is lower in energy and at this instance energy starts being transferred from the thermal reservoir to the spin system. The process is illustrated in Fig.~\ref{fig:splitprocess} for the excitation peak $\mathbf{q}=0.5\mathbf{e}_z(2\pi /a)$. The transfer of energy from the thermodynamic reservoir to the magnetic system leads to an increasing peak size. This initial energy transfer to the magnetic system leads to a reduction of the size of the average magnetization. However as the switching process continues the orientation of the average moment changes, reducing the angle between magnetization and the external field. This changes the relative strengths of the three peaks in the split. As the angle is reduced below 90 degrees the positive Zeeman peak becomes largest implying an in average positive shift of the excitation. This reverses the energy transfer between the magnetic system and the thermal reservoir. Energy is now being transported out of the magnetic system leading to reduced peak sizes and an increased saturation magnetization. Within the atomistic model there are changes in the magnitude of the macrospin during the switching process. This does however not affect the precessional frequency of the macro moment which is only dependent on the size of the external field. 

\begin{figure}
\includegraphics*[width=0.45\textwidth]{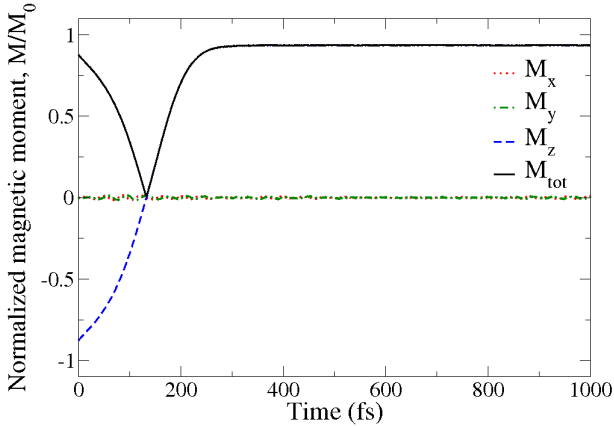}
\caption{(Color online) Magnetic switching process in a 1000~T constant external field applied anti-parallel with respect to the magnetization at 300K.}
\label{fig:T300H1000AP}
\end{figure}

\begin{figure}
\includegraphics*[width=0.45\textwidth]{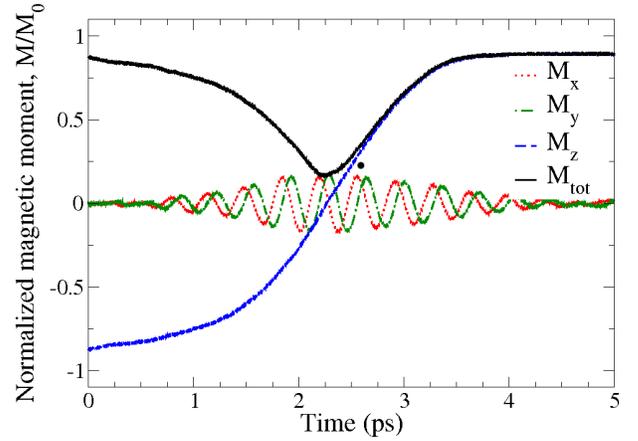}
\caption{(Color online) Magnetic switching process in a 100~T constant external field applied anti-parallel with respect to the magnetization at 300K}
\label{fig:T300H100AP}
\end{figure}

If the initial angle between the external field and the magnetization is increased toward anti-parallel alignment, the effect of shrinking of the magnetization during the switching process is enhanced. This is illustrated in Fig.~\ref{fig:T300H1000AP} where a 1000~T anti-parallel field is applied to the magnetization where the shrinking is total and accounts for the whole switching process. Note that at a certain point in the process the size of the macroscpin (sum of all atomic spins) is zero, after which it grows back to a saturation value again. For weaker external fields the effect is reduced. This is illustrated in Fig.~\ref{fig:T300H100AP} where a 100~T external field is applied anti-parallel to the magnetization and although the size of the applied field is much smaller, the magnitude of the magnetization is still heavily effected by the external field.

To summarize this section, by simulations of magnetization dynamics of atomic resolution, we have explored magnetic switching in the limit of large external fields. A significant difference from LLG theory is seen for large fields. The results may aid understanding ultrafast switching experiments with ultra large field pulses.

\section{Conclusions}

A full account of the details of an atomic spin dynamics method has been given. 
A comprehensive description of all pertinent details for performing spin dynamics simulations, such as equations of motions, models for including temperature, damping mechanisms, methods of extracting data and numerical schemes for performing the simulations, have been presented.
Various ways to analyze spin dynamics simulations have been presented including spin-correlations. 
The developed method can be applied in a first principles mode, where all interatomic exchange is calculated self-consistently, or it can be applied with frozen parameters estimated from experiments or calculated for a fixed spin-configuration. 

Applications of the method have been made to several systems, with interatomic exchange calculated from first principles, primarily for bcc Fe from a frozen spin-configuration. Various switching phenomena have been studied, such as the dynamics of a spin-system when the easy axis rapidly changes direction. Furthermore, we have simulated the spin dynamics of a system with
a large applied field parallel, antiparallel and at an angle to the macro spin (the sum of all atomic spins). In this particular system we show that the macrospin model breaks down. This happens when the applied field is of similar size as the interatomic exchange. An experimental realization of this is possibly best obtained for nano-structured magnetic multilayers, since the interatomic exchange interaction (which in many systems influences the magnetic properties heavily) typically is of the same size as magnetic field available in the laboratory.

\acknowledgments
Financial support from the Swedish Foundation for Strategic Research (SSF), Swedish Research Council (VR), the Royal Swedish Acadamy of Sciences (KVA), Liljewalchs resestipendium and Wallenbergstiftelsen is acknowledged. Calculations have been performed at the Swedish national computer centers UPPMAX, HPC2N and NSC.

\appendix

\section{Langevin spin dynamics}
The Fokker-Planck equation describes the time evolution of a non-equilibrium probability distribution. The Fokker-Planck equation corresponding to the stochastic Landau Lifshitz Gilbert (LLG) equation has been derived.\cite{Brown63} \label{sec:appb}

The general form of the Langevin equations can be written as
\begin{equation}
\frac{dy}{dt}=A_i(y,t)+\sum_{k} B_{ik}(y,t)L_k(t),
\label{eq:langevingen}
\end{equation}
where
\begin{equation}
<L_k(t)>=0, \quad <L_k(t)L_l(s)>=2D\delta_{kl}\delta(t-s).
\end{equation}
The stochastic LLG equation can be written in the general form of a Langevin-equation by identifying the coefficients
\begin{equation} 
A_i=\gamma {[\mathbf{m} \times \mathbf{B}_{\textrm{eff}}- \frac{\alpha}{\mathnormal{m}}\mathbf{m} \times (m \times \mathbf{B}_{\textrm{eff}})]}_i,
\label{eq:adim}
\end{equation} 
\begin{equation} 
B_{ik}=\gamma[\sum_{j} \epsilon_{ijk}m_j + \frac{\alpha}{m}(m^2\delta_{ik}-m_im_k) ].
\label{eq:bdim}
\end{equation} 
The time evolution of the general form, using Stratanovich calculus is given by
\begin{eqnarray}
\frac{\partial P}{\partial t}= - \sum_{i} \frac{\partial}{\partial y_i}[(A_i+D\sum_{jk}B_{jk}\frac{\partial B_{ik}}{\partial y_j})P]+\nonumber \\ +\sum_{ij} \frac{\partial^2}{\partial y_i \partial y_j}[(D\sum_{k} B_{ik}B_{jk})P]
\end{eqnarray}
We then arrive at the following Fokker-Plank equation for the time evolution of the probability distribution, $P(\mathbf{m})$, of the atomic moments, 
\begin{eqnarray}
\frac{\partial P}{\partial t}=-\frac{\partial}{\partial \mathbf{m}}\{[ \gamma \mathbf{m} \times \mathbf{B}_{\textrm{eff}} - \gamma \frac{\alpha}{m}\mathbf{m} \times (\mathbf{m} \times \mathbf{B}_{\textrm{eff}}) + \nonumber \\ 
+ {\frac{1}{2\tau} \mathbf{m} \times (\mathbf{m} \times \frac{\partial}{\partial \mathbf{m}})]} P\},
\label{eq:sllgfp}
\end{eqnarray}
where we have defined $\tau$ as,
\begin{equation}
\frac{1}{\tau}=2D\gamma^2(1+\alpha^2)
\label{eq:tnllg}
\end{equation}
The Fokker-Planck equation associated with the stochastic LLG equation must satisfy the correct thermal-equilibrium properties. In thermal equilibrium $P(\mathbf{m})$ must have the form of the Boltzmann distribution,
\begin{equation}
P_0(\mathbf{m}) \propto \textrm{exp}[-\beta \mathscr{H}(\mathbf{m})].
\label{eq:boltzman}
\end{equation}
This condition on the Fokker-Planck equation is not consistent with $\mathrm{\hat{I}}$to calculus. For Stratanovich calculus, we can find a condition on $\tau$ that makes the equations fulfill the equilibrium requirement. First note that 
\begin{equation} 
\mathbf{B}_{\textrm{eff}} = -\frac{\partial \mathscr{H}}{\partial \mathbf{m}}.
\end{equation}
Using Eq.~(\ref{eq:boltzman}) we can write 
\begin{equation}
\frac{\partial P_0}{\partial \mathbf{m}} = \beta {\mathbf{B}_{\textrm{eff}}} P_0.
\end{equation}
Hence, the first term of Eq.~(\ref{eq:sllgfp}), 
\begin{equation}
\frac{\partial}{\partial \mathbf{m}}{(\gamma \mathbf{m} \times {\mathbf{B}}_{\textrm{eff}})} P_0 
\end{equation}
vanishes and the remainder can be written as
\begin{eqnarray}
\frac{\partial P}{\partial t}=-\frac{\partial}{\partial \mathbf{m}}\{[ - \gamma \frac{\alpha}{\mathnormal{m}}\mathbf{m} \times (\mathbf{m} \times \mathbf{B}_{\textrm{eff}}) + \nonumber \\ + {\frac{\beta}{2\tau_{N}} \mathbf{m} \times (\mathbf{m} \times \mathbf{B}_{\textrm{eff}})]} P\},
\end{eqnarray}
From this we see that a requirement for a stationary solution, or $\partial P / \partial t=0$, is that $\gamma \alpha / m = \beta/2 \tau_N$. Hence, resulting in
\begin{equation}
\tau_N = \frac{1}{\alpha} \frac{m}{2 \gamma k_B T}.
\end{equation} 
In Eq.~(\ref{eq:tnllg}) we see that the temperature determines the amplitude of the random field, $\mathbf{b}_i$, and this is how temperature enters our simulations. The amplitudes are finally given by,
\begin{equation}
D = \frac{\alpha}{1+\alpha^2} \frac{k_B T}{\gamma m},
\label{eq:dllg2}
\end{equation} 





\end{document}